# ASTRONOMY IN THE CHURCH:
## THE CLEMENTINE SUNDIAL IN SANTA MARIA DEGLI ANGELI, ROME

**Prof. Costantino Sigismondi, University of Rome "La Sapienza"**  sigismondi@icra.it
www.santamariadegliangeliroma.it

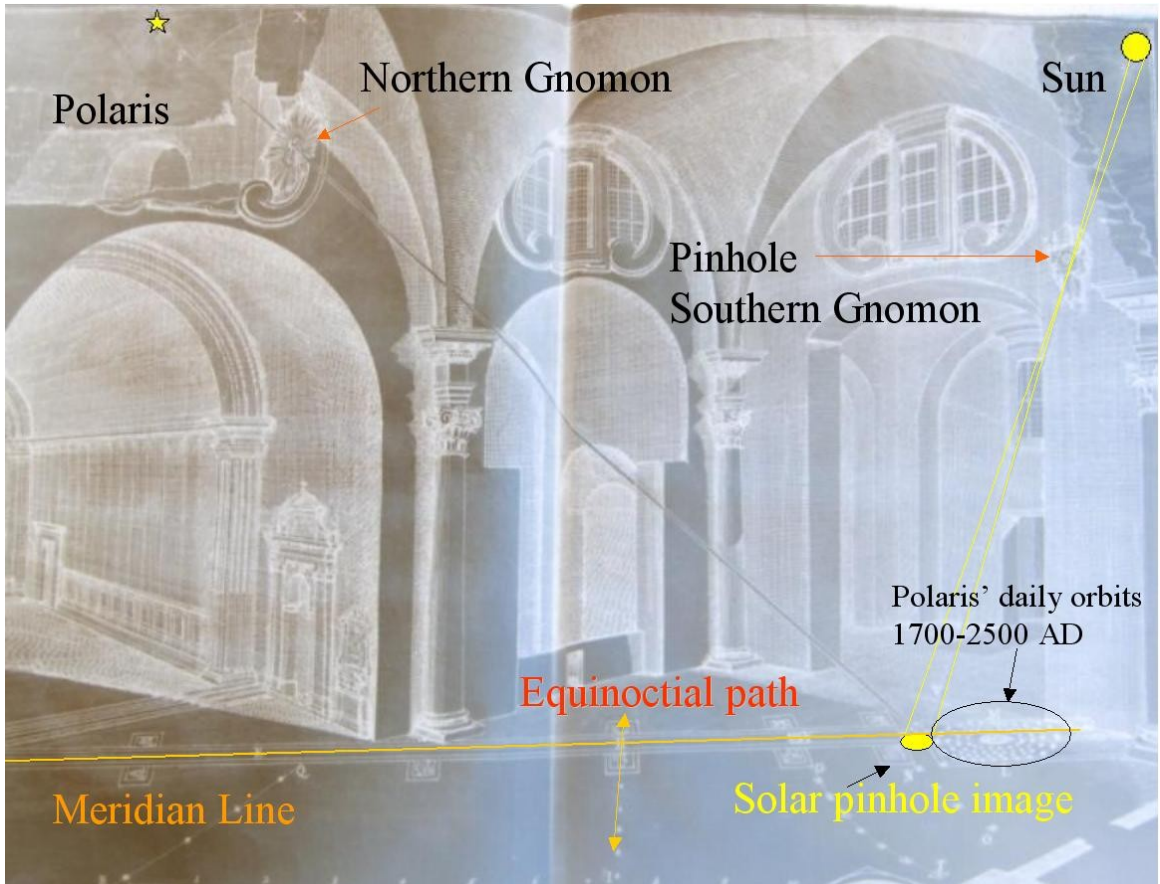

## HISTORY

Pope Clement XI (1700-1721) ordered Francesco Bianchini (1662-1729) to build a Meridian Line. Bianchini was the Secretary of the Commission for the Calendar. He chose the Basilica of Santa Maria degli Angeli because of the stability of its roman walls and foundations and its suitable dimension. The Line built by Giandomenico Cassini (1625-1712) in 1655 in San Petronio, Bologna, was the model for Bianchini, who improved it by allowing the observation of stellar transits.

Pope Clement XI inaugurated the Great Sundial on October 6, 1702, the annual Feast of St. Bruno, founder of the Carthusian Order, whose statue is in the entrance of the Church. Carthusian monks ruled this Basilica for three centuries until 1884.

## SCIENCE WITH THE CLEMENTINE GNOMON

Stability over centuries of the ancient walls where the pinhole is located is a requirement for making high precision astrometry, such as the measurement of the inclination of the Earth axis over its orbit plan.

The exact durations of mean lunar month and of tropical year were other scientific tasks of this Meridian Line, motivated also by civil and religious exigencies. The parameters introduced by the Gregorian reformation (1582) of the Calendar have been tested with this instrument in the first years of work. Daily solar transits gave the exact time for Angelus prayer at noon; the instant of vernal equinox, to which Easter is linked, was also computable with this Line. Using the Northern Gnomon, Bianchini determined the apparent elevation of the Celestial Pole within 1" of accuracy, the difference from the true latitude was discovered only three decades later, due to stellar aberration.

## Observation of Stars

In the 18th century it was possible to open the window holding the southern pinhole, and, even in daylight, stellar transits were recorded and precisely timed with pendulum mechanical clocks. The accuracy of such clocks was better than 1 s per day, and the observations of stellar transits allowed their synchronization with sidereal time. The names of some bright stars are engraved on the meridian marbles in the positions that they had in 1702. Sirius, the brightest star, is near the number 161.

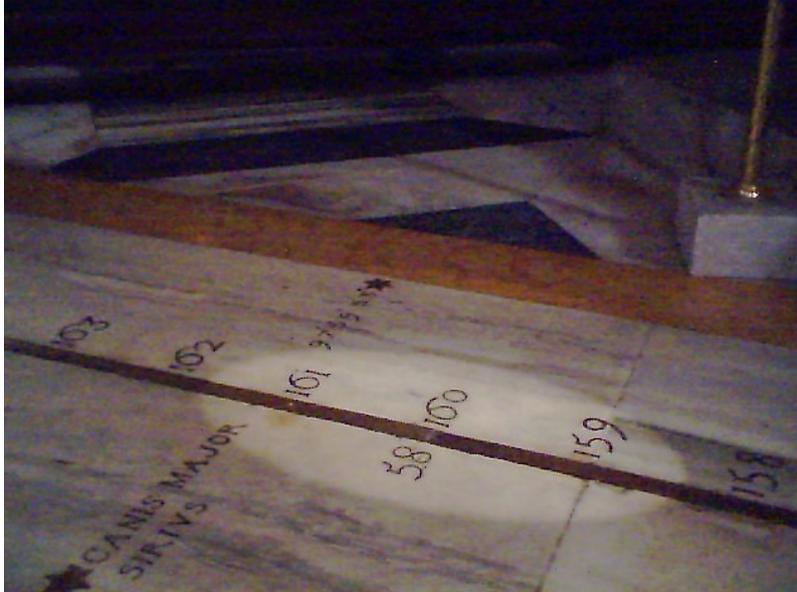

**Sun transits at the same altitude of Sirius as in 1702 (Feb. 4, 2005)**

Bianchini reported the observations of Sirius at noon in June-July 1703. The windows of the Basilica were darkened with external tents and, at the same time when the pinhole solar image was crossing the floor like nowadays, the star was observed from the telescope. The prolongation of the line of sight of the telescope on the Line corresponded to the altitude of the star at the meridian transit. Other stars have their name and right ascension engraved on the marbles, being the meridian altitude indicated by the position of the brass star.

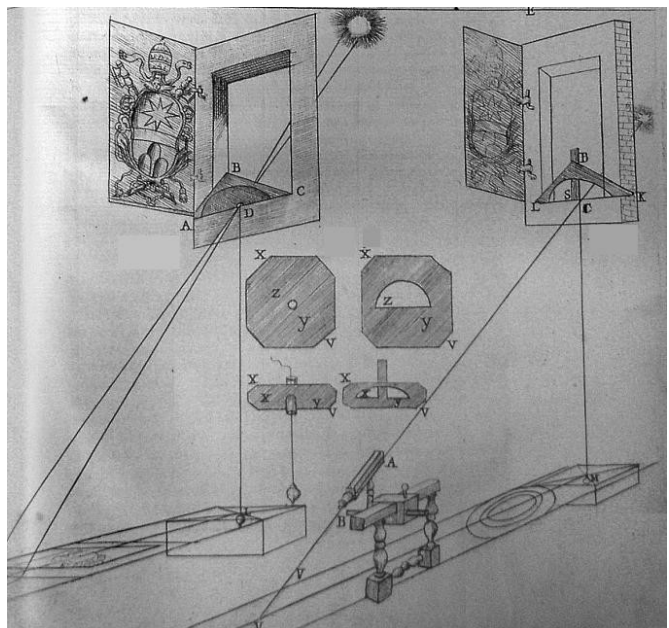

**The telescope used in the observations of stars on the Clementine Line**

# Equinoxes and Solstices: tropical year and inclination of the ecliptic

This *hybrid feature* of the Clementine Gnomon to measure solar and stellar transits allowed Bianchini to accomplish in 1703 the whole measurement of the duration of the tropical year, which was usually made by comparing observations very widely spread in time. The time difference between the transit of a fixed star and that one of the Sun produced the ecliptic longitude of the Sun, which has to be 0°, 90°, 180° and 270° respectively for spring equinox, summer solstice, fall equinox and winter solstice. Evenly spaced sectors of 30° longitudes are traditionally linked to the names of **Zodiac Signs**, for example: Aries stands for 0° to 30° in ecliptic longitude and so on. Because of equinoxes precession zodiac signs are no longer linked with corresponding constellations, but only to the ecliptic longitudes of the Sun as in all classical astronomy. The measurement of the altitude of the Sun above the horizon, which attains an extreme of the range of all possible values, was also useful to indicate the actual value of the inclination of Earth's axis to the ecliptic plane, once corrected the measurement for the refraction effect. For this purpose Bianchini used the tables of refraction computed by Cassini in San Petronio, Bologna.

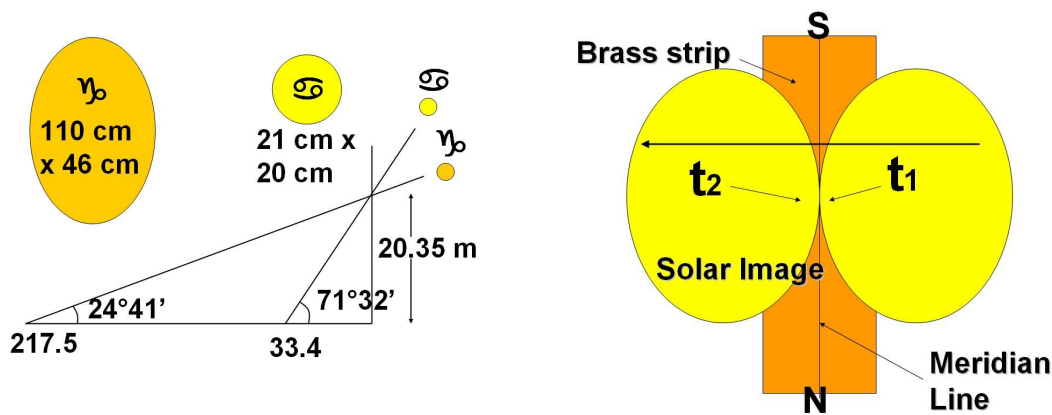

**Solar images at solstices, and their altitudes. Timings of the transit on the Line.**

## The solar image

The solar image is produced by the pinhole at 20.34 m in height; it is an ellipse and its minor axis is perpendicular to Line and measures ~1/100 of the distance from the pinhole. The image is reversed according to the optical laws of *camera obscura*. Air turbulence produce a continuous vibrations of the whole image at high frequency visible even to the naked eye. This phenomenon generates an intrinsic uncertainty in the transit's timing of ±0.4 s. Using a videocamera and ten lines disposed parallely to the main Line, with the display of a radio-controlled reference clock in the same video, we can average these timings improving statistically the measurement of the time of transit up to ±0.1 s. With this parallel transits method it has been possible to verify along the years the slow drift of transit time with respect to the ephemerides, due to the progressive slowing down of Earth's rotation: one leap second has been necessary to be added to the Universal Time Coordinated (UTC) in order to keep it close to the ephemerides. The small deviation of the Line from true North of ~ 4'30" Eastwards has been measured with this technique, comparing the delays of transits at both solstices with respect to ephemerides.

## Visual Timings

The accuracy of visual timings is ± 0.5 s and the transit time is obtained averaging t1 and t2, as represented in the figure, according to the Bianchini's method.

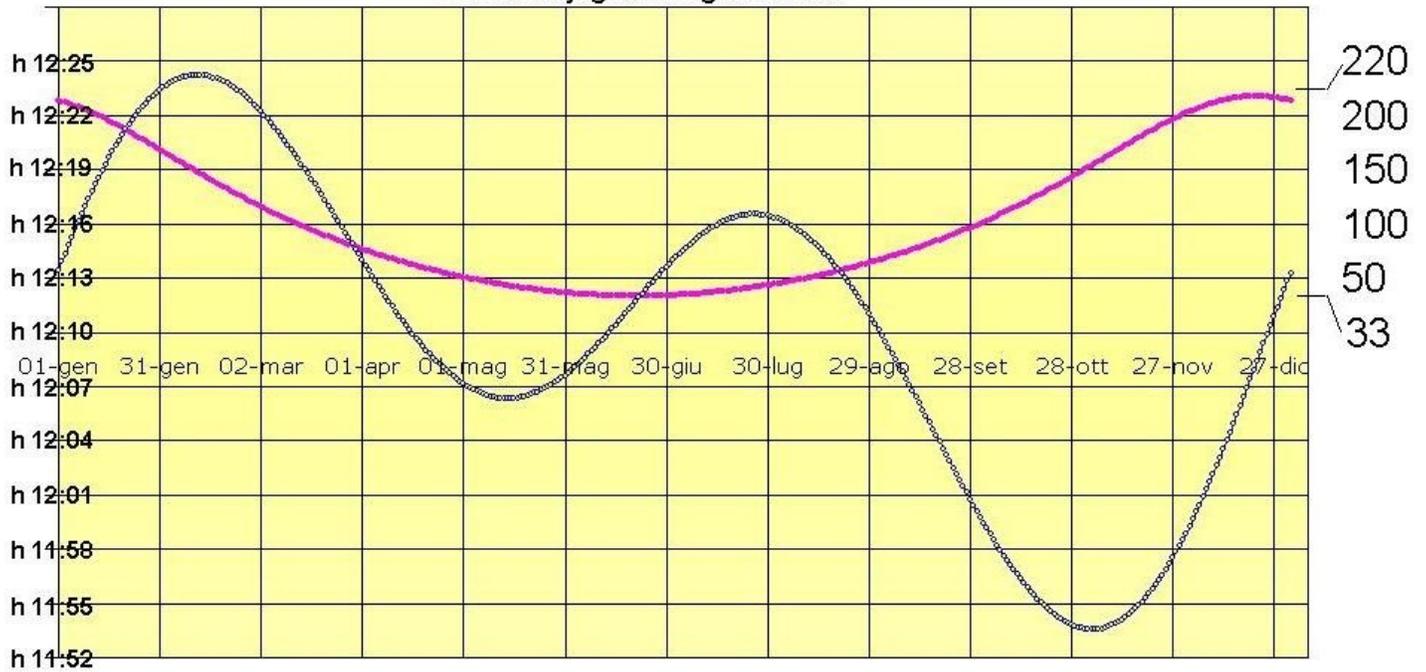

## HOW TO READ THE MERIDIAN LINE

Following this timetable it is possible to know times of transit and their locations on the Line. Locations are expressed in *Centesimae Partes* of the pinhole's height, and are indicated on the line from 33 to 220. Each of those numbers are related to the zenithal distance $z$ of the Sun at the meridian by the formula *CP=100·tan(z)*. $z$ is also written in degrees on the west side of the Line: for example 78≈100·tan(38°) in the **photo of March 29, 2006 solar eclipse.**

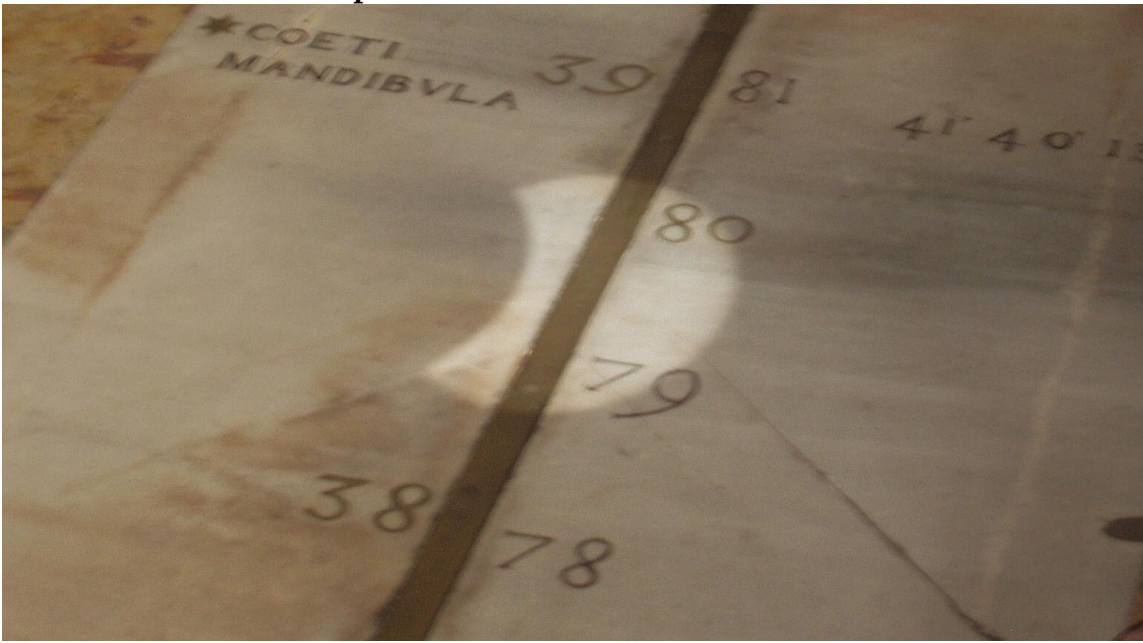